\documentclass{aa}
\authorrunning{R. Landman et al.}

\usepackage{graphicx}
\usepackage[varg]{txfonts}
\usepackage[normalem]{ulem}
\begin{document}

   \title{Detection of OH in the ultra-hot Jupiter WASP-76b}

   \subtitle{}
   \author{R. Landman \inst{1},
          A. S\'{a}nchez-L\'{o}pez \inst{1},
          P. Molli\`{e}re\inst{2},
          A.Y. Kesseli \inst{1},
          A.J. Louca \inst{1},
          \and
          I.A.G. Snellen \inst{1}
          }

   \institute{Leiden Observatory, Leiden University, Postbus 9513, 2300 RA Leiden, The Netherlands\\
    \email{rlandman@strw.leidenuniv.nl}
         \and
             Max-Planck-Institut für Astronomie, Königstuhl 17, 69117 Heidelberg, Germany
             }

   \date{Received XXX; accepted YYY}

 
  \abstract
   {Ultra-hot Jupiters have dayside temperatures at which most molecules are expected to thermally dissociate. The dissociation of water vapour results in the production of the hydroxyl radical (OH). While OH absorption is easily observed in near-infrared spectra of M dwarfs, which have similar effective temperatures as ultra-hot Jupiters, it is often not considered when studying the atmospheres of ultra-hot Jupiters. Ground-based high-resolution spectroscopy during the primary transit is a powerful tool for detecting molecular absorption in these planets.}
   {We aim to assess the presence and detectability of OH in the atmosphere of the ultra-hot Jupiter WASP-76b.}
   {We use high-resolution spectroscopic observations of a transit of WASP-76b obtained using CARMENES. After validating the OH line list, we generate model transit spectra of WASP-76b with petitRADTRANS. The data are corrected for stellar and telluric contamination and cross-correlated with the model spectra. After combining all cross-correlation functions from the transit, a detection map is constructed. Constraints on the planet properties from the OH absorption are obtained from a Markov chain Monte Carlo analysis.}
   {OH is detected in the atmosphere of WASP-76b with a peak signal-to-noise ratio of 6.1. From the retrieval we obtain $K_p=232 \pm 12$ km/s and a blueshift of $-13.2 \pm 1.6$ km/s, which are offset from the expected velocities. Considering the fast spin rotation of the planet, the blueshift is best explained with the signal predominantly originating from the evening terminator and the presence of a strong dayside-to-nightside wind. The increased $K_p$ over its expected value (196.5 km/s) is, however, a bit puzzling. The signal is found to be broad, with a full width at half maximum of $16.8^{+4.6}_{-4.0}$ km/s. The retrieval results in a weak constraint on the mean temperature of 2700-3700 K at the pressure range of the OH signal.}
   {We show that OH is readily observable in the transit spectra of ultra-hot Jupiters. Studying this molecule can provide insights into the molecular dissociation processes in the atmospheres of such planets.}

    
    \maketitle

%

\section{Introduction}
Ultra-hot Jupiters (UHJs) are gas giant planets that orbit extremely close to their host star; as such, they receive large amounts of radiation, resulting in dayside temperatures of >2200 K \citep{Parmentier2018A&A_UHJ_Wasp121}. They provide a unique opportunity to study atmospheric chemistry under extreme conditions that are not present in any of the planets in the Solar System. Furthermore, because of their short orbital periods and bloated atmospheres, they are well suited for atmospheric characterisation through transmission spectroscopy. At these high dayside temperatures, most molecules are thermally dissociated into their atomic constituents. This has led to a large range of detections of atoms and ions in the atmospheres of UHJs \citep[e.g.][]{Hoeijmakers2019A&A_Kelt9, Cassasayas_Barris2019A&A_Atoms_Mascara2b, Hoeijmakers2020A&A_wasp121b,Yan_2019A&ACaII_wasp33, Tabernero2021A&A_wasp76}. However, some diatomic and triatomic molecules are still expected to be present at temperatures typical for UHJs. For example, CO, $\textrm{H}_2$O, OH, FeH, TiO, and VO absorption can be observed in the spectra of M dwarfs \citep{Reiners2018_Carmenes_survey, Rajpurohit2018A&A_Mdwarf}, which have similar effective temperatures as the daysides of UHJs, albeit at higher surface gravities. Furthermore, recombination of these molecules is expected to occur on the limbs or nightside of the planet \citep{Parmentier2018A&A_UHJ_Wasp121}. However, high spectral resolution searches for these molecules in UHJs have mostly been unsuccessful \citep[e.g.][]{Kesseli2020_FeH, Merritt2020A&A_nondetection_TiO_wasp121}. A detection of TiO was claimed in the atmosphere of WASP-33b \citep{Nughroho2017AJ_TiO}, but not reproduced by \citet{Herman2020AJ_TiO} or \citet{Serindag2021A&A_TiO_wasp33}. Recently, new evidence for TiO in the dayside spectrum of WASP-33b has been presented \citep{Cont2021arXiv_TIO_W33}.

Space-based low spectral resolution observations of UHJs with the Wide Field Camera 3 (WFC3) of the Hubble Space Telescope (HST) also show mostly featureless and blackbody-like emission spectra \citep[e.g.][]{Mansfield2018AJ_HAT7b_WFC3, Kreidberg2018AJ_wasp103_HST, Arcangeli2018ApJ_dissociation}.
This has been suggested as being the result of the dissociation of water on the dayside of these planets, in combination with the consequent increase in H$^-$ opacity \citep{Lothringer2018ApJ_UHJ, Parmentier2018A&A_UHJ_Wasp121, Arcangeli2018ApJ_dissociation,  Gandhi2020AJ_H-}. Indeed, \citet{Parmentier2018A&A_UHJ_Wasp121} show that for a subset of UHJs, which have sufficiently high dayside temperatures and sufficiently low surface gravities, it is expected that water is partially or completely thermally dissociated. In the case of WASP-76b, their models show that about half of the water vapour is thermally dissociated at the expected 1.4 $\mu$m photosphere \citep[][Fig. 13]{Parmentier2018A&A_UHJ_Wasp121}.

One of the products of the dissociation of water is the hydroxyl radical (OH). Still, OH is often not considered when studying atmospheres of UHJs. There have been some attempts at modelling the expected OH abundance in hot Jupiters and rocky exoplanets \citep{Moses2011ApJ_disequilibrium_chemistry, Hu2012ApJ_photochemistry_rocky, Miguel2014ApJ_disequilibrium}. These have mostly focused on planets with equilibrium temperatures of $<2000$ K. For these planets, the main OH production mechanism is the photolysis of water, which enhances its abundance in the upper atmosphere over the equilibrium case. Indeed, OH has been detected in the atmospheres of the Earth \citep{Meinel1950ApJ_OH_earth}, Venus \citep{Piccioni2008A&A_OH_venus}, and Mars \citep{Clancy2013Icar_OH_Mars}. However, in the case of UHJs, thermal dissociation of water is expected to be the main OH production mechanism, increasing its abundance and making it potentially easier to detect in these planets. \citet{Nugroho2021_OH} presented the first evidence for OH in the dayside emission spectrum of WASP-33b at a significance of 5.5$\sigma$, showing that this is indeed the case. In contrast, water was only marginally detected in WASP-33b, indicating that it is largely thermally dissociated in the upper atmosphere.

WASP-76b is an UHJ orbiting an F7V star with an orbital period of 1.8 days and an equilibrium temperature of $\sim2160 \pm 40$ K \citep{West2016A&A_wasp76}. Observations of the transmission and emission spectrum with HST/WFC3 revealed the presence of water in the atmosphere of WASP-76b \citep{Tsiaras2018AJ_survey_hst_wfc3_wasp76, Edwards2020AJ_hst_wfc3_wasp76}. Furthermore, the WFC3 observations showed hints of TiO and a thermal inversion in its atmosphere \citep{Edwards2020AJ_hst_wfc3_wasp76}. High-resolution transmission spectroscopy from the ground using HARPS showed the presence of sodium \citep{Seidel2019A&A_sodium_wasp76, Zak2019AJ_wasp76_harps_sodium}, which was later confirmed using the HST's Space Telescope Imaging Spectrograph \citep{VonEssen2020A&A_sodium_wasp76_hst_stis}. Furthermore, Na I, Fe I, Li I, Mg I, Ca II, and Mn I were detected using ESPRESSO by \citet{Tabernero2021A&A_wasp76}. However, no molecular absorption from species such as TiO or VO was observed in the optical. \citet{Ehrenreich2020_iron} showed that the iron absorption feature in the ESPRESSO data is asymmetric during the transit, which was later confirmed with HARPS-N \citep{Kesseli2021ApJ_Fe_asymmetric_wasp76}. The iron absorption was found to increase in strength along the transit and to progressively blueshift from between 0 and -5 km/s during ingress up to $-11$ km/s at the end of the transit. The tidally locked rotation of the planet, with a rotation speed of  $\sim5.3$ km/s, results in the absorption on the trailing limb being more blueshifted than the absorption of the leading limb. Furthermore, winds blowing from the hot dayside to the cooler nightside can result in an additional blueshift of several km/s. The blueshift of $-11 \pm 0.7$ km/s of the iron signal during the second half of the transit can thus be explained by the combination of the signal mostly originating from the trailing evening-side limb and the presence of a strong dayside-to-nightside wind. The absence of the signal on the morning side was interpreted as iron condensing across the nightside. This was the first evidence for a chemical gradient in the atmosphere of an exoplanet. Three-dimensional Monte Carlo radiative transfer simulations by \citet{Wardenier2021arXiv_wasp76} show that condensation of iron on the nightside could indeed explain the observations. However, they can also be explained by a large temperature contrast between the leading and trailing limbs of the planet. \\

In this paper we report on the detection of OH in the atmosphere of WASP-76b using high-resolution transmission spectroscopy with CARMENES (Calar Alto high-Resolution search for M dwarfs with Exoearths with Near-infrared and optical Echelle Spectrographs). This paper is structured as follows: Section \ref{sec:obs} reports on the observations and data processing. Section \ref{sec:models} describes the modelling of the OH transmission spectrum of WASP-76b. Section \ref{sec:results} presents the results, which are then discussed in Sect. \ref{sec:discussion}.

\section{Observations and data reduction}\label{sec:obs}
\subsection{Observations}
We used archival data obtained with CARMENES, a high-resolution spectrograph at the 3.5 meter telescope at the Calar Alto Observatory. It covers both the near-infrared (NIR) and the visible with spectral resolutions of $R\sim 80,400$ and $R\sim 94,600$, respectively \citep{Quirrenbach2014_carmenes_instrument, Quirrenbach2018SPIE_CARMENES}. We downloaded the publicly available NIR observations of the 4 October 2018 transit of WASP-76b from the CAHA Archive\footnote{http://caha.sdc.cab.inta-csic.es/calto/jsp/searchform.jsp}. The retrieved CARMENES spectra had already been reduced using the CARACAL pipeline \citep{Zechmeister2014A&A_data_reduction,Caballero2016_caracal}. The observations consist of 44 exposures of 498 seconds each, with an average signal-to-noise ratio (S/N) per pixel of $\sim$56 in the continuum. We excluded the last eight and the first out-of-transit spectra from our analysis because of a lower S/N. The remaining 35 spectra cover orbital phases between -0.068 and +0.053. The relevant parameters of WASP-76b and its host star are given in Table \ref{tab:parameters_wasp76}. Throughout this paper, we only use spectral orders of the NIR channel that are not dominated by telluric absorption, as shown in Fig. \ref{fig:model}. 

\begin{table}[htbp]
\label{tab:parameters_wasp76}
\def\arraystretch{1.3}
\caption{Parameters of WASP 76b and its host star.}
\begin{tabular}{l|c}

\hline
Parameter                             & Value                                         \\ \hline
Stellar Spectral Type                & F7                                           \\
Stellar radius, $R_*$                  & $1.756 \pm 0.071 R_\odot$                      \\
Systemic velocity, $v_{sys}$             & $-1.11\pm 0.5$ km/s                          \\
Period, $P$                            & $1.80988198^{+0.00000064}_{-0.00000056}$ days     \\
Planet radius, $R_p$                   & $1.856^{+0.077}_{-0.076} R_{\textrm{jup}}$                 \\
RV semi-amplitude, $K_p$               & $196.52 \pm 0.94$ km/s                             \\
Mid-transit time (BJD), $T_c$                & $58080.626165^{+0.000418}_{-0.000367}$           \\
Planet surface gravity, $g_p$          & $6.4 \pm 0.5$ m/s$^{2}$                  \\       \hline
\end{tabular}
\vspace{1ex}
{\raggedright \textbf{References}: All values are obtained from \citet{Ehrenreich2020_iron} \par}
\end{table}

\begin{figure*}[ht]
    \centering
    \includegraphics[width=0.8\linewidth]{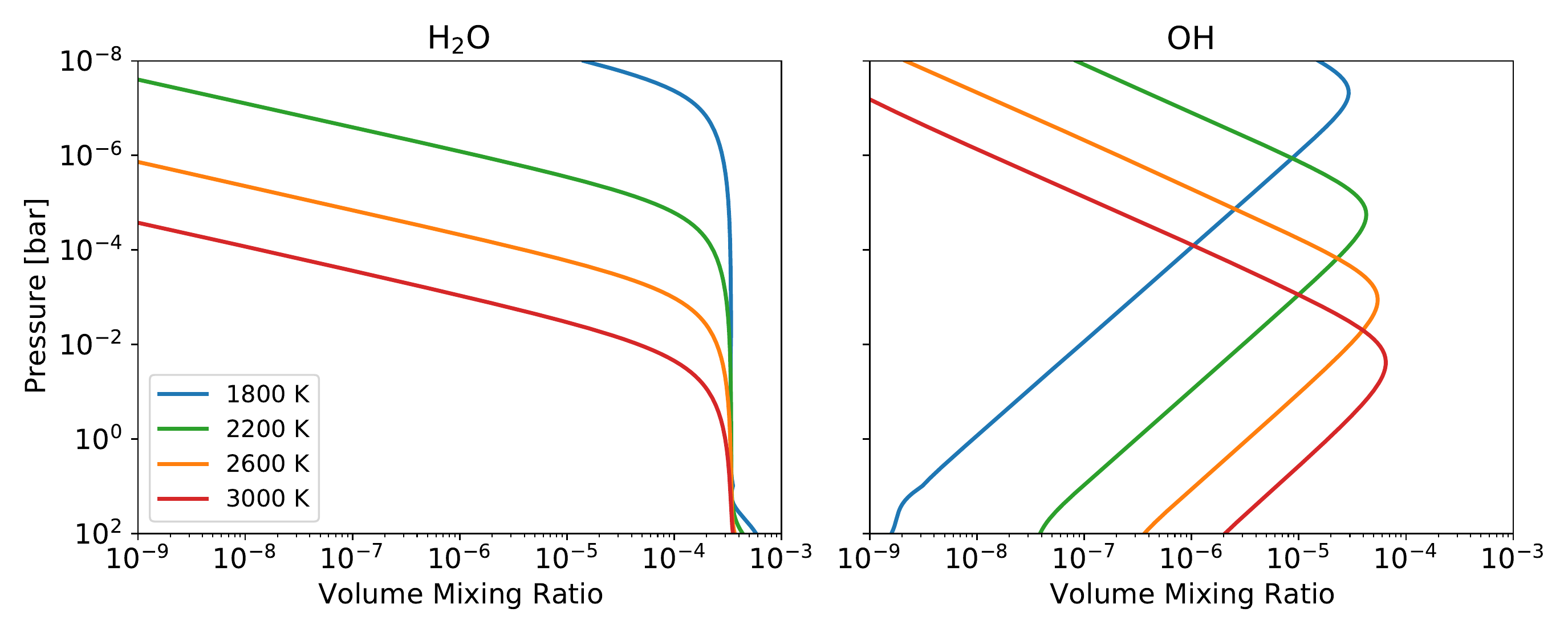}
    \caption{Volume mixing ratio of H$_2$O and OH using the chemical equilibrium model from \citet{Molliere2017A&A_chemical} and an isothermal P-T profile for different temperatures.}
    \label{fig:abundance}
\end{figure*}

\subsection{Stellar and telluric corrections}
 We followed a similar cleaning procedure as previous work with CARMENES data \citep[e.g.][]{Alonso_floriano2019_waterhd189, Sanchez_lopez2019_waterhd209, Kesseli2020_FeH}. All these steps were performed separately for each of the used orders. First, we removed the continuum by fitting a third-order polynomial and dividing by this fit. Subsequently, we flagged wavelength bins that have on average less than 40\% flux or 10\% excess flux with respect to the continuum. This removes those parts of the spectra that are dominated by telluric absorption or sky emission lines. Finally, we flagged any remaining $5\sigma$ outliers. The flagged data points are not considered in the remainder of the analysis.\\
 
The spectra are still dominated by telluric and stellar features at this point. We used the SysRem algorithm \citep{Tamuz2005MNRAS_sysrem} to correct for this. SysRem uses an iterative, noise-weighted principal component analysis to remove any shared systematics between light curves. SysRem has been successfully used for removing tellurics in high-resolution spectroscopy data in previous work \citep[e.g.][]{Birkby2017AJ_water_51Pegb, Nughroho2017AJ_TiO, Alonso_floriano2019_waterhd189, Sanchez_lopez2019_waterhd209}. Such a blind detrending algorithm can be applied because the radial velocity (RV) of the planet changes by tens of km/s during the transit, while stellar and telluric features virtually remain at the same wavelength location. However, when too many SysRem iterations are applied, this leads to the subtraction of the planet signal itself. On the other hand, removing too few modes leads to residual tellurics, which increases the noise and can lead to spurious signals. We used nine SysRem iterations for each of the spectral orders and models in this paper. Other studies have optimised the number of SysRem iterations for each of the spectral orders separately \citep[e.g.][]{Sanchez_lopez2019_waterhd209,SanchezLopez2020A&A_hazy}. This optimisation requires an accurate model of the strength of the planet signal to estimate when self-subtraction of the planet signal occurs, which is not trivial. Therefore, we chose a conservative approach and used the same number of SysRem iterations for each of the spectral orders and models. Appendix \ref{app:sysrem} shows the impact of changing the number of SysRem iterations on the S/N of both the observed signal and an injected signal. The pixel uncertainties used for SysRem are obtained from the result of the CARACAL pipeline and are propagated through the preprocessing steps.
 
 \subsection{Cross-correlation}
After removing the stellar and telluric features, individual spectral lines from the planet are still hidden in the noise. The signal from all of the tens to hundreds of individual spectral lines can be combined by cross-correlating with a representative template spectrum of the targeted species. The cross-correlation technique for detecting molecular absorption or emission from exoplanet atmospheres has been extensively used \citep[e.g.][]{Snellen2010Nature_HD209, Brogi2012Nature_taubootis, Snellen2014Nature_betapic, Giacobbe2021Natur_molecular_absorption_hd209}. This cross-correlation is performed by linearly interpolating the template spectrum to the Doppler-shifted wavelengths of the data and taking the inner product. We did this for all orders simultaneously with an RV range from -300 km/s to +300 km/s with a step size of 1.3 km/s. This was done for each of the spectra, which results in a 2D cross-correlation map. Since the radial component of the orbital velocity of the planet changes as a function of the orbital phase, the planet signal would appear as a slanted line in this cross-correlation map (e.g. as indicated in Fig. \ref{fig:ccf_map}). The RV of the planet is given by:
\begin{equation}\label{eq:rv}
    RV = v_{bar}+v_{sys} + K_p \sin (2\pi\phi).
\end{equation}
Here, $v_{bar}$ is the barycentric velocity of the Earth at the time of the exposure, $v_{\textrm{sys}}$  the systemic velocity of the system, $K_p$ the RV semi-amplitude of the planet, and $\phi$ the orbital phase of the planet. The cross-correlation functions (CCFs) were then shifted to the planet's rest frame for each exposure. To further enhance the signal, we then summed the CCFs from all in-transit exposures, obtaining a total CCF for the entire transit. In contrast to some previous studies, we did not employ a weighting scheme for different spectral orders or exposures. The S/N of the cross-correlation peak was then estimated by subtracting the mean value and dividing by the standard deviation. The mean and standard deviation were calculated excluding the central region of $\pm 50$ km/s.

\section{Models}\label{sec:models}
The cross-correlation requires a model of the OH transmission spectrum of the planet. These model spectra are generated using the high-resolution mode of petitRADTRANS \citep{Molliere2019_prt}, a radiative transfer code developed specifically for the characterisation of exoplanet atmospheres. As input it requires pre-computed opacity grids of the desired molecules, the abundance profiles of the molecules, the pressure-temperature (P-T) profile of the atmosphere, and the surface gravity and radius of the planet. The OH line list was retrieved from the HITEMP database \citep{Rothman2010_HITEMP}. This line list was validated on the spectrum of the M dwarf GX And, and was found to be sufficiently accurate for this study (see Appendix \ref{app:line_val}). 

We also included continuum opacity due to $\textrm{H}_2-\textrm{H}_2$, $\textrm{H}_2-\textrm{He}$ scattering and H$^-$ in our models. Since high-resolution transmission spectroscopy is not very sensitive to absolute abundances and changes in the P-T profile, we simply used an isothermal atmosphere. The OH abundance and the mean molecular weight were determined using the chemical equilibrium model from \citet{Molliere2017A&A_chemical}, assuming solar abundance ratios and metallicity. The OH abundances for isothermal atmospheres of different temperatures are shown in Fig. \ref{fig:abundance}. It shows that for higher temperatures, OH is expected to be located lower in the atmosphere with a higher volume mixing ratio. This is because at a higher pressure, water vapour thermally dissociates at a higher temperature, as shown in Fig. \ref{fig:abundance}. 

\begin{figure}
    \centering
    \includegraphics[width=0.95\linewidth]{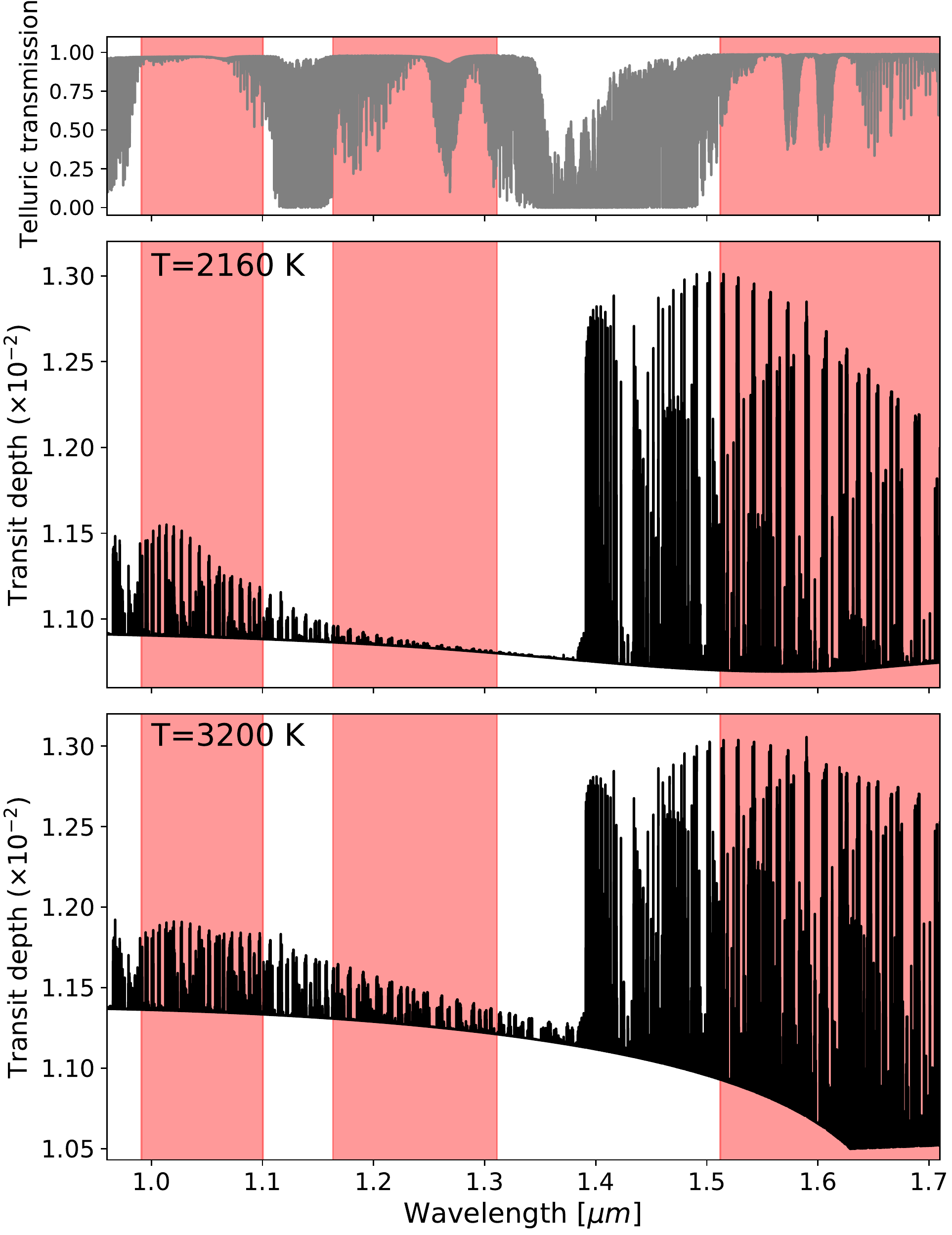}
    \caption{\textbf{Top:} Telluric transmission as a function of wavelength obtained from ESO SkyCalc \citep{Noll2012A&A_skycalc}. \textbf{Middle:} Transit depth as a function of wavelength, modelled using
    petitRADTRANS over the entire CARMENES NIR channel and assuming an isothermal atmosphere at the planet equilibrium temperature of 2160 K. The red region indicates the spectral range used for our analysis. \textbf{Bottom:} Same but for T=3200 K.
    }
    \label{fig:model}
\end{figure}

The output of petitRADTRANS consists of the effective transit radius of the planet at each sampled wavelength. The transit depth for an isothermal atmosphere at the equilibrium temperature of 2160 K is shown in Fig. \ref{fig:model}. This shows that there is some absorption around 1 $\mu m$, but most of the OH opacity is located between 1.4 $\mu m$ and 1.7 $\mu m$. This is in agreement with our findings from the line validation using an M dwarf (Appendix \ref{app:line_val}). We included all spectral orders that are not heavily contaminated by tellurics, as indicated in Fig. \ref{fig:model}.

The resulting models were subsequently convolved with a Gaussian kernel that matched the spectral resolution of CARMENES. Furthermore, the RV change of the planet during a single exposure of 498 seconds is about 3.9 km/s. This results in a further smearing of the signal. To take this into account, we convolved the model with a box car with this width. Finally, we removed the continuum of the model by fitting a third-order polynomial and divided by this.

\section{Results}\label{sec:results}
\begin{figure}[htbp]
    \centering
    \includegraphics[width=\linewidth]{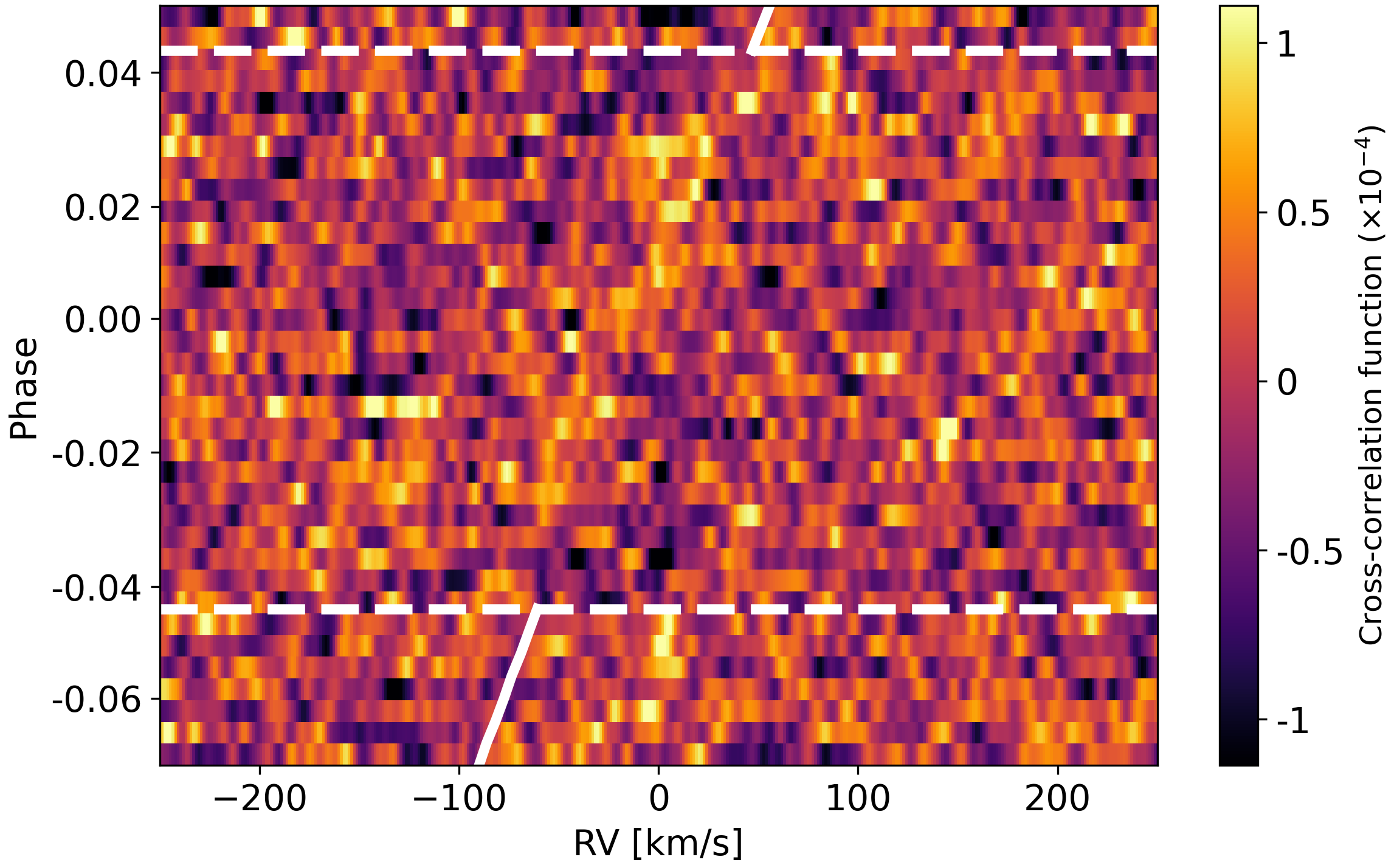}
    \caption{Resulting CCF for each of the spectra, with the RV in the Earth's rest frame on the horizontal axis and the planet orbital phase on the vertical axis. The horizontal dashed white lines indicate the start and end of the transit, while the slanted vertical lines show the expected slope and location of the planetary signal}.
    \label{fig:ccf_map}
\end{figure}

\begin{figure}[htbp]
    \centering
    \includegraphics[width=\linewidth]{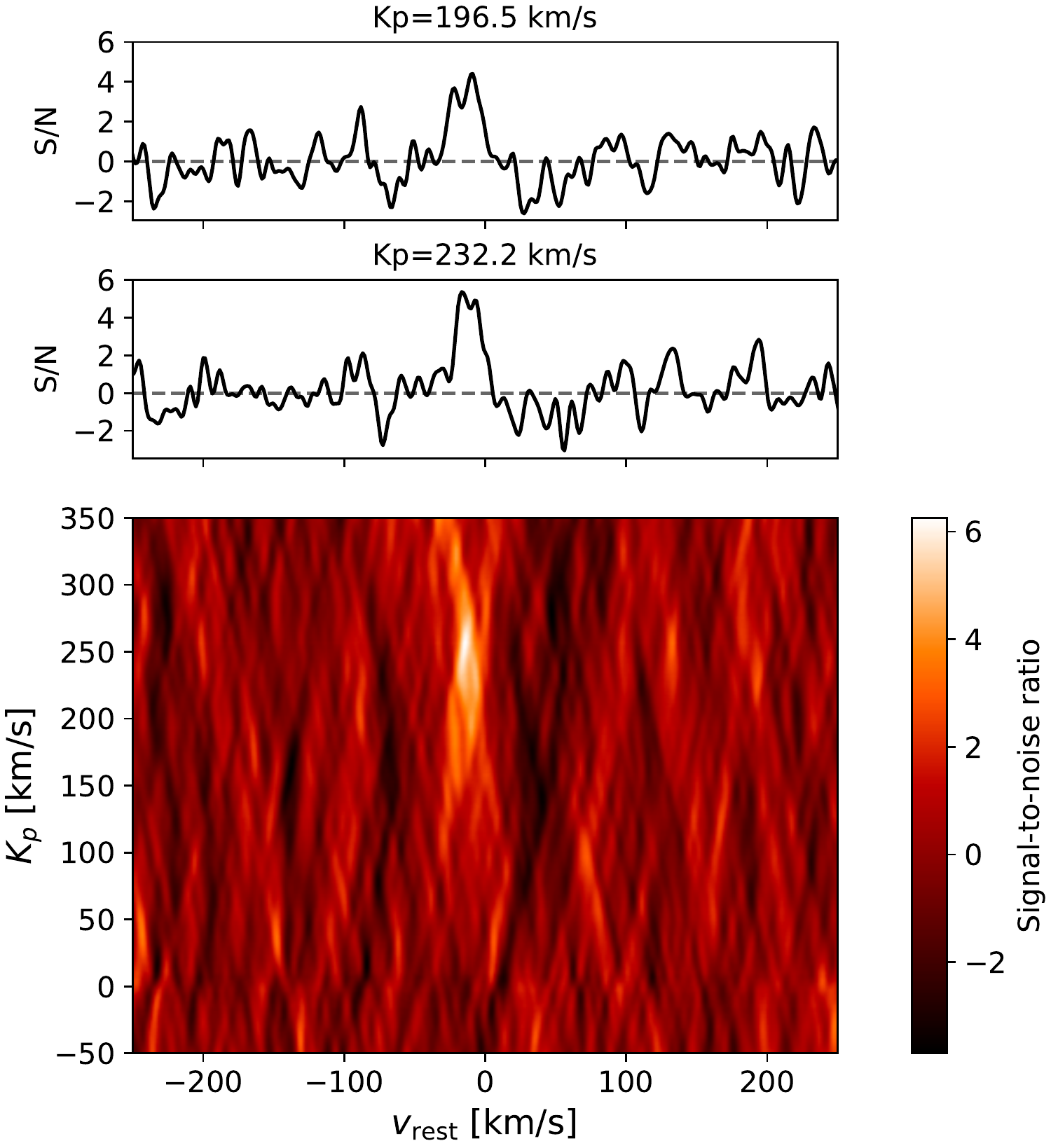}
    \caption{\textbf{Bottom}: Detection significance as a function of RV semi-amplitude, $K_p$, and velocity relative to the planet rest frame, $v_{\textrm{rest}}$, showing a $\sim6 \sigma$ peak. \textbf{Top}: Detection significance at the expected $K_p$ of 196.5 km/s and at $K_p=232.2$ km/s, which is obtained from the retrieval.}
    \label{fig:Kp_vsys}
\end{figure}

The cross-correlation map we obtained using the model that assumes an isothermal atmosphere with a temperature of 2160 K is shown in Fig. \ref{fig:ccf_map}. The slanted white lines indicate where the planet signal is expected to be. This cross-correlation map was converted to a detection significance for each $K_p$-$v_{\textrm{rest}}$ combination by co-adding the CCFs in the rest frame of the planet following Eq. \ref{eq:rv}. The resulting signal-to-noise map is shown in Fig. \ref{fig:Kp_vsys}. We detect OH with a peak S/N of 6.1. Also shown are the 1D CCFs for both $K_p$ = 196.5 km/s, the expected value, and 232.2 km/s, which is the value we obtain from the retrieval.

We performed a Markov chain Monte Carlo (MCMC) retrieval to constrain the atmospheric conditions in the terminator region of WASP 76b, following the log-likelihood mapping from \citet{Gibson2020MNRAS_retrieval}:
\begin{equation}
    \ln \mathcal{L} = -\frac{N}{2}\ln \left[\frac{1}{N} \sum_{i=1}^N\frac{(f_i - \alpha m_i)^2}{\sigma_i^2} \right]
.\end{equation}
Here, $N$ is the number of wavelength bins, $\alpha$ the model scaling parameter, $f_i$ the data, $m_i$ the model, and $\sigma_i$ the error at wavelength bin $i$. As free parameters we used $K_p$, the RV of the OH signal in the planet rest frame, $v_{\textrm{rest}}$, the temperature in the planet's atmosphere, the full width at half maximum (FWHM) of the Gaussian broadening kernel applied to the transmission spectrum obtained from petitRADTRANS, and the scaling parameter $\alpha$. We used uniform priors with bounds specified in Table \ref{tab:priors} and an isothermal P-T profile, and calculated the OH abundance following the chemical equilibrium model from \citet{Molliere2017A&A_chemical}. To decrease the computation time, we pre-computed the models on a grid with a temperature resolution of 50 K and interpolated linearly between the two closest temperatures. The reason that we included the FWHM of the signal is because this can point to dynamics in the planet's atmosphere. Broadening of the signal can, for example, be caused by the rotation of the planet \citep{Snellen2014Nature_betapic, Brogi2016ApJ_rotation_winds_hd189}, equatorial super-rotation, or high-altitude vertical winds \citep{Seidel2020A&A_winds}. Indeed, absorption features from WASP-76b have been observed to be significantly broadened \citep{Seidel2019A&A_sodium_wasp76, Tabernero2021A&A_wasp76}.

\begin{figure*}[htbp]
    \centering
    \includegraphics[width=0.8\linewidth]{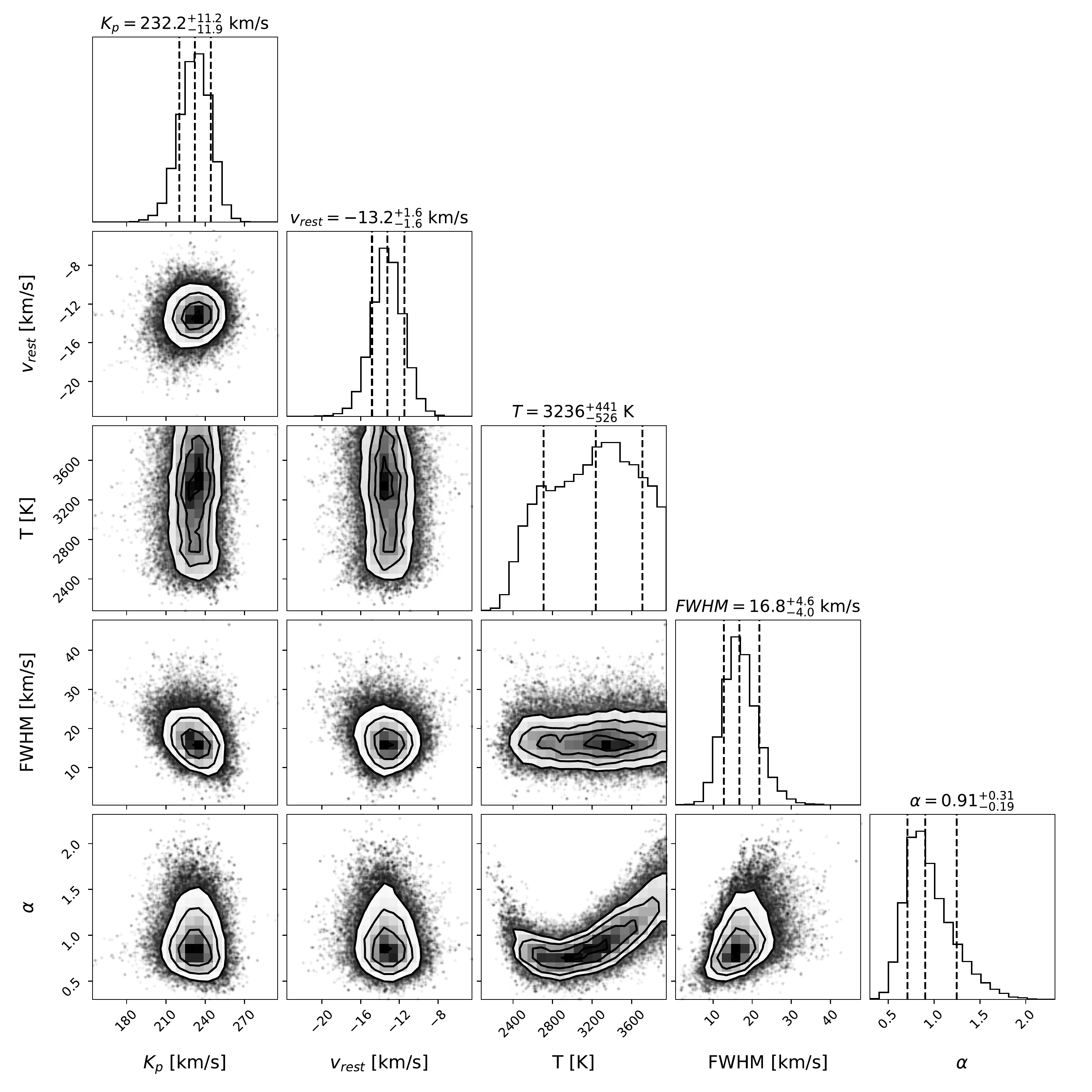}
    \caption{Resulting posterior distributions for the OH signal in WASP-76b obtained from the MCMC retrieval. The dashed vertical lines indicate the 16th, 50th, and 84th percentiles.}
    \label{fig:corner}
\end{figure*}

The retrieval was done using the \textit{emcee} package for python \citep{Foreman_Mackey2013PASP_emcee}. We used 50 walkers, ran the MCMC for 3000 iterations, and discarded the first 1000 iterations as the burn-in. The resulting corner plot of our retrieval is shown in Fig. \ref{fig:corner}. \\

\begin{table}[htbp]
\label{tab:priors}
\def\arraystretch{1.3}
\caption{Parameters included in the retrieval and their priors.}
\begin{tabular}{l|c}

\hline
Parameter                             & Prior bound                                        \\ \hline
Radial velocity semi-amplitude, $K_p$ & 100 - 300 km/s \\
Planet rest velocity, $v_{\textrm{rest}}$ & -50 - 50 km/s \\
Planet temperature, $T$ &  1500 - 4000 K \\
Full width at half maximum, $FWHM$ & 0 - 50 km/s \\
Scaling parameter, $\alpha$ & 0 - 5 
\end{tabular}
\vspace{1ex}

\end{table}

\section{Discussion}\label{sec:discussion}
From the retrieval we obtain $K_p=232.2\pm 12$ km/s; this is offset from the expected planet orbital velocity of 196.5 km/s, which was calculated from the stellar mass, inclination, and orbital period \citep{Ehrenreich2020_iron}. From the posterior distribution we find that the expected $K_p$ is excluded at a significance of $2.7\sigma$. Furthermore, we find that the signal has a very significant blueshift of $-13.2\pm1.6$ km/s with respect to the planet rest frame.

\citet{Wardenier2021arXiv_wasp76} show that rotation and atmospheric dynamics can lead to peaks away from the expected $K_p$ and $v_{\textrm{rest}}$. The signal from the evening limb is blueshifted due to a combination of planet rotation and a strong dayside-to-nightside wind, leading to a total blueshift of $-11 \pm 0.7$ km/s \citep{Ehrenreich2020_iron}. On the other hand, on the morning limb, the planet rotation and dayside-to-nightside wind are in opposite directions, resulting in a signal between 0 and -5 km/s. Our retrieved $v_{\textrm{rest}}$ is similar to that found for iron at the end of the transit \citep{Ehrenreich2020_iron}. We therefore argue that the OH signal presented here is dominated by the evening terminator. This may be the result of a chemical asymmetry as the atmosphere at the evening terminator consists of gas emerging from the hotter dayside, where the OH abundance is expected to be higher (see Fig. \ref{fig:abundance}). In addition, photochemical processes may have had the chance to further enrich the atmosphere in OH \citep{Moses2011ApJ_disequilibrium_chemistry, Miguel2014ApJ_disequilibrium}. Alternatively, this may be caused by the larger scale height on the evening terminator or clouds forming on the morning terminator, which both lead to the evening terminator dominating the transmission signal \citep{Wardenier2021arXiv_wasp76, Savel2021_wasp76_no_iron_rain}. However, we cannot rule out contributions from the morning terminator at velocities between 0 and -5 km/s, especially given the broadness of the signal.

The offset in $K_p$, if real, is a bit puzzling. The simulations from \citet{Wardenier2021arXiv_wasp76} indicate that 3D effects would mostly lead to lower $K_p$ values than expected. This means that effects not included in their model need to be invoked to explain the OH signal obtained here. The lower $K_p$ values simulated by \citet{Wardenier2021arXiv_wasp76} result from an interplay between planet rotation, variations in viewing angle, and a particular iron-distribution and/or thermal structure in the planet atmosphere. This results in the morning-side and evening-side terminators dominating the transmission spectrum in the first and second half of the transit, respectively, leading to a decrease in the observed Doppler shift of the signal and thus a lower $K_p$. The marginally higher-than-expected $K_p$ from the planet orbit observed here would imply that the blueshift of the signal decreases along the transit. While the origin of this is unclear, it is interesting that \citet{SanchezLopez2021_wasp76_molecules} find similar increased $K_p$ values for H$_2$O and HCN. Furthermore, \citet{Kesseli2021_wasp76_chemical_inventory} also find evidence for increased $K_p$ values for specific species (Li, Na, and H). Higher S/N observations will be needed to validate the offset in $K_p$ observed here and to study the time-resolved velocity shift of the OH signal in more detail. Furthermore, 3D global circulation models of the OH distribution and the subsequent transmission spectrum will be needed to interpret the signals, but this is beyond the scope of this work.

We find a weak constraint on the mean atmospheric temperature of 2700-3700 K (68\% confidence interval). While the error margins are large, this is much higher than the equilibrium temperature of $2160 \pm 40$ K \citep{West2016A&A_wasp76} and is similar to the value obtained in \citet{Seidel2021_verticalwinds} for Na absorption. This supports the evidence that the signal originates from the hotter parts of the planet. However, a shared retrieval with both low- and high-resolution spectra may provide better constraints on the thermal structure in the atmosphere \citep{Brogi2017ApJ_LRS+HRS}.

We find a value of $16.8^{+4.6}_{-4.0}$ km/s for the FWHM of the signal. This means the absorption features are significantly broader than expected from instrumental and observational effects only. A broadened absorption signal in WASP-76b was also found for iron \citep{Ehrenreich2020_iron}, sodium \citep{Seidel2019A&A_sodium_wasp76}, and many other species \citep{Tabernero2021A&A_wasp76}. \citet{Seidel2021_verticalwinds} show that this broadening may be explained by including high velocity vertical winds in the upper atmosphere. The broad OH signal observed here may again be evidence for strong dynamics in the upper atmosphere of WASP-76b. The scaling parameter $\alpha=0.91^{+0.31}_{-0.19}$ indicates that the strength of the observed signal is close to the expected strength from the models.

An important caveat to the retrieval is that we used models that assume a simple 1D isothermal atmosphere, but including 3D effects is necessary for interpreting the transmission spectra of UHJs \citep[e.g.][]{ Caldas2019A&A_biased_retrieval,MacDonald2020ApJ_biased_retrieval,Wardenier2021arXiv_wasp76}. Furthermore, the procedure of removing the stellar and telluric contamination using SysRem may alter the planetary signal and bias the obtained posteriors \citep{BrogiLine2019AJ_retrieval}. It may especially lead to a decrease in the signal strength, decreasing the scaling parameter, and a smearing of the signal, increasing the observed FWHM.

Finally, our detection shows that OH is one of most abundant spectroscopically active molecules in the atmosphere of WASP-76b and that it should be considered when studying similar UHJs. Our detection reinforces the conclusions made in \citet{Nugroho2021_OH}, who detect OH in the dayside emission spectrum of WASP-33b. WASP-33b has a dayside temperature even higher than that of WASP-76b, meaning that water is expected to thermally dissociate at higher pressures as compared to WASP-76b. Since emission spectroscopy generally probes higher pressures than transmission spectroscopy, this is consistent with both detections. OH may therefore be used as a probe for the thermal dissociation processes in these UHJs.
 
\begin{acknowledgements}
R.L., A.S.L, A.K, I.S. acknowledge funding from the European Research Council (ERC) under the European Union's Horizon 2020 research and innovation program under grant agreement No 694513. P.M. acknowledges support from the European Research Council under the European Union's Horizon 2020 research and innovation program under grant agreement No. 832428. This research has made use of the Spanish Virtual Observatory (http://svo.cab.inta-csic.es) supported by the MINECO/FEDER through grant AyA2017-84089.7.

\end{acknowledgements}

\bibliographystyle{aa}
\bibliography{references}

\begin{appendix}

\section{Validation of the OH line list}\label{app:line_val}
\begin{figure}[htbp]
    \centering
    \includegraphics[width=\linewidth]{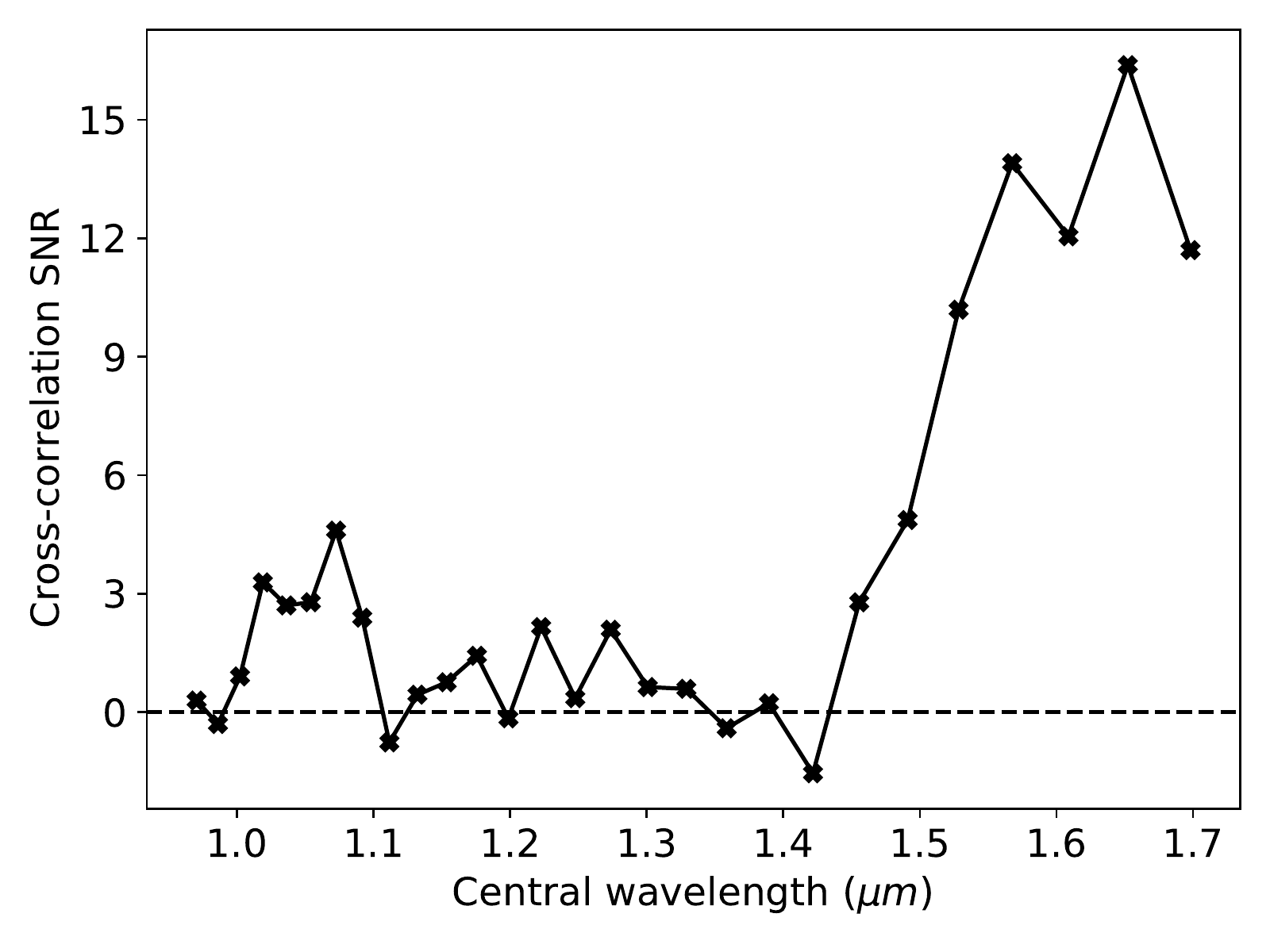}
    \caption{Cross-correlation S/N with the spectrum of GX And with the OH model template for each of the spectral orders from the CARMENES NIR channel.}
    \label{fig:line_validation}
\end{figure}

We validated the HITEMP OH line list \citep{Rothman2010_HITEMP} on the spectra of an M dwarf. We used CARMENES observations of GX And that were published as part of a survey searching for exoplanets around M dwarfs \citep{Reiners2018_Carmenes_survey}. GX And is an M1 star with a $J$-band magnitude of 5.25. We used the observation taken on 10 November 2016, the OH lines of which are already denoted in the appendix of \citet{Reiners2018_Carmenes_survey}. The cross-correlation template was generated using petitRADTRANS \citep{Molliere2019_prt}, as described in Sect. \ref{sec:models}. Figure \ref{fig:line_validation} shows the resulting cross-correlation S/N at the star's systemic velocity for each of the spectral orders. We see that the highest S/N is obtained in the reddest orders of the CARMENES data.

\section{Impact of SysRem iterations on the signal-to-noise ratio}\label{app:sysrem}
\begin{figure}[htbp]
    \centering
    \includegraphics[width=\linewidth]{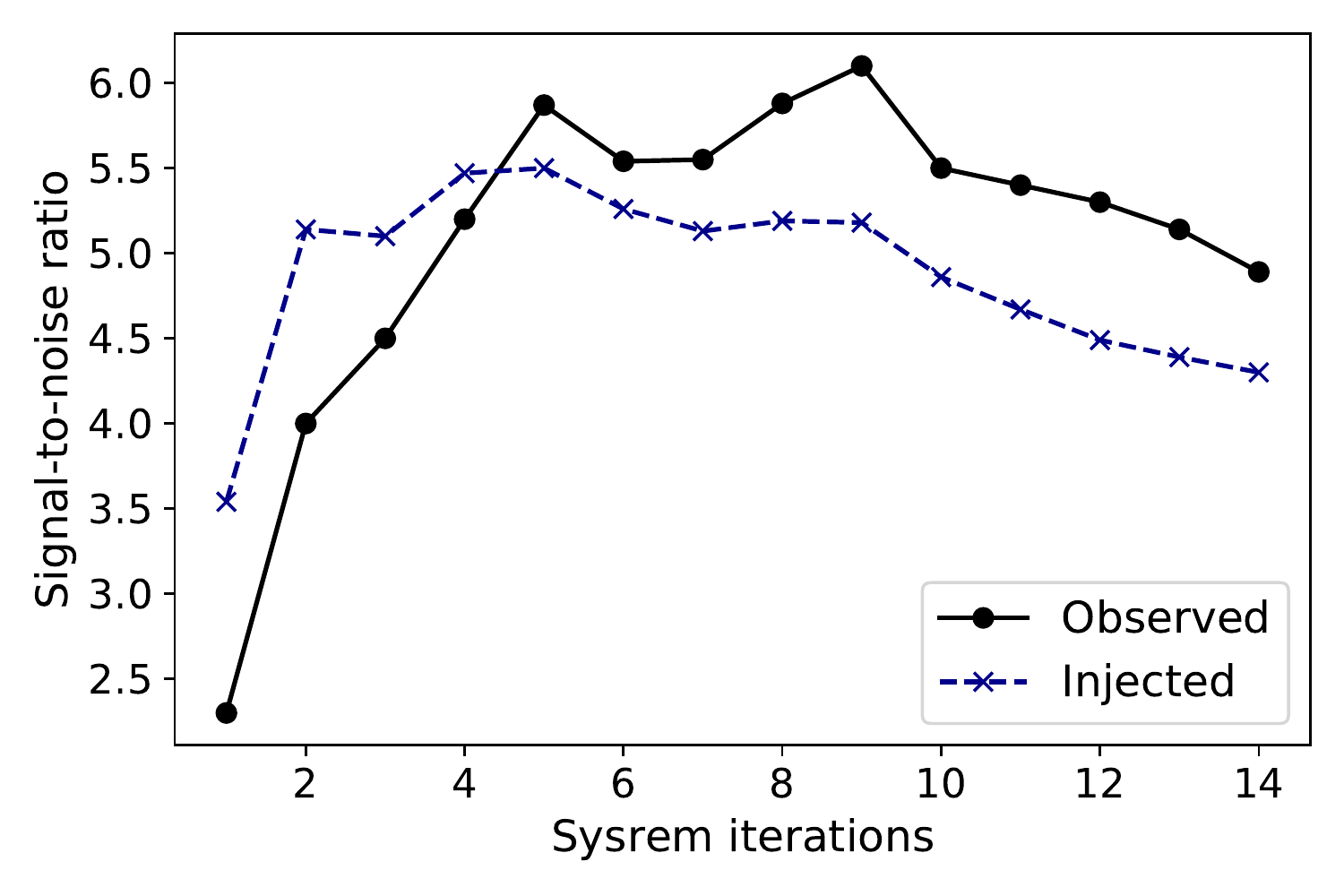}
    \caption{Obtained peak S/N as a function of the number of SysRem iterations applied to the data.}
    \label{fig:sysrem_iterations}
\end{figure}

Figure \ref{fig:sysrem_iterations} shows the effect of changing the number of SysRem iterations on the peak S/N of the observed OH signal. Also shown is the retrieved S/N for an injected signal at the expected strength with an offset of +100 km/s. While the maximum S/N for the injected signal is obtained at five SysRem iterations, we observed some clear residuals in the cross-correlation map due to telluric contamination. We therefore chose to use nine SysRem iterations, which gives the highest S/N of the observed signal and gives a much cleaner CCF map.
\end{appendix}

\end{document}